\documentclass[nocopyrightspace]{sig-alternate-10pt}
\usepackage{graphicx}
\usepackage{amsmath}
\usepackage{amsfonts}
\usepackage{amssymb}
\usepackage{eucal}
\usepackage{subfigure}
\usepackage{multirow}
\usepackage{dblfloatfix}
\usepackage[colorlinks=true,pdfstartview=FitV,linkcolor=black,citecolor=black,urlcolor=blue,plainpages=false]{hyperref}
\usepackage[numbers,sort&compress]{natbib}
\graphicspath{{figs//}}

\begin{document}

\def\newblock{\hskip .11em plus .33em minus .07em} 

\def\sharedaffiliation{%
\end{tabular}
\begin{tabular}{c}}

\title{Pushing the limits of Full-duplex: Design and Real-time Implementation}


\numberofauthors{1} \author{ \alignauthor{Achaleshwar Sahai, Gaurav
    Patel and Ashutosh
    Sabharwal}\\ \email{\normalsize{\{as27,gpatel,ashu\}@rice.edu}}
  \sharedaffiliation \affaddr{Department of Electrical and Computer
    Engineering} \\ \affaddr{Rice University } \\ \affaddr{Technical Report TREE1104} }

\maketitle

\begin{abstract}
Recent work has shown the feasibility of single-channel full-duplex
wireless physical layer, allowing nodes to send and receive in the
same frequency band at the same time. In this report, we first design
and implement a real-time 64-subcarrier 10~MHz full-duplex OFDM
physical layer, FD-PHY. The proposed FD-PHY not only allows
synchronous full-duplex transmissions but also selective asynchronous
full-duplex modes. Further, we show that in over-the-air experiments
using optimal antenna placement on actual devices, the
self-interference can be suppressed upto 80dB, which is 10dB more than
prior reported results. Then we propose a full-duplex MAC protocol,
FD-MAC, which builds on IEEE~802.11 with three new mechanisms --
shared random backoff, header snooping and virtual backoffs. The new
mechanisms allow FD-MAC to discover and exploit full-duplex
opportunities in a distributed manner. Our over-the-air tests show
over 70\% throughput gains from using full-duplex over half-duplex in
realistically used cases.
\end{abstract}

\section{Introduction}

Half-duplex communication, where a node can either transmit or receive
in a single channel, is the commonly imposed constraint in the design
of all practical wireless networks.  In the last two decades, many
works~\cite{CWRadar2,DivisionFreeDuplex,MultibandFrontEnd,RaghavanCollocated,FD_Margetts,FDMicrosoft,Choi:2010aa,Duarte:2010aa}
have reported experiments and/or models for full-duplex
communications. Perhaps the most encouraging results were reported by
two groups simultaneously~\cite{Choi:2010aa,Duarte:2010aa} which used
off-the-shelf hardware to demonstrate that single-channel full-duplex
wireless can in fact be implemented and provides measurable gains over
half-duplex systems. However, most work till date has limited its
attention to two nodes exchanging information with each other, with
the focus on physical layer feasibility -- a crucial first
step. However, there is no prior work on the design of medium access
protocols which leverage full-duplex communications in multi-node
networks.

In this report, we propose the first full-duplex random access
protocol, FD-MAC. Further, we implement a real-time OFDM-based
full-duplex physical layer (FD-PHY for short) and the proposed FD-MAC
on a WARP-based testbed. Our major contributions for FD-PHY and FD-MAC
are as follows.

{\bf FD-PHY}: We develop and implement a real-time full-duplex capable
physical layer, FD-PHY. The OFDM-based FD-PHY has 64 sub-carriers and
occupies 10~MHz bandwidth. The key challenge in full-duplex
communications is the large self-interference caused by a node's own
transmissions, which can completely swamp the packets from other
nodes. Thus, analog cancellation (passive and/or active) is essential
to reduce the power of self-interference compared to the packet of
interest \emph{before} the analog-to-digital converter converts the
signal from the antenna. We implement an active analog cancellation
which injects an appropriately scaled canceling signal at the receive
antenna, to reduce the self-interference. This active cancellation is
implemented on a per subcarrier basis, and can thus be applied to any
OFDM PHY with arbitrary number of subcarriers.

In~\cite{Choi:2010aa,Duarte:2010aa}, the self-interference
cancellation was performed both in analog and in digital baseband,
together providing nearly 70~dB of attenuation to self-interference
signal. We explore another avenue of attenuating the self-interference
-- the role of physical placement and orientation of transmit and
receive antennas on actual mobile devices, like laptops and
tablets. We conduct extensive experiments by mounting the antennas on
an iPad-sized device for different antenna configurations. The main
finding is that device-induced attenuations combined with analog
cancellation can lead to 80~dB of self-interference suppression, even
\emph{without} baseband cancellation.  This finding further
strengthens the case for actual deployment of full-duplex
communication in mobile devices.

Further, the proposed FD-PHY is also capable of enabling
\emph{asynchronous} full-duplex communications in some cases, which
further expands the design space for medium access protocols. We show
that a full-duplex capable node can begin to receive a packet from a
node while transmitting to another node, albeit with a 3~dB loss for
the same bit error rate (BER). However, the other case where a
full-duplex-capable node wants to transmit a packet while receiving a
packet is not possible to implement reliably. This imposes important
constraints on medium access protocols, which the proposed FD-MAC
completely adheres to.

{\bf FD-MAC}: Leveraging the capabilities of FD-PHY, we develop and
implement a random access protocol FD-MAC for infrastructure-based
WiFi-like networks, where all flows are between an access point and
mobile units. We use IEEE~802.11 packet structure with an additional
FD header.  The key challenge in maximally using full-duplex
capability is to discover the opportunities to send and receive at the
same time in a completely distributed manner. Since nodes only have
the knowledge about the packets in their own queues, discovering a
full-duplex opportunity requires sharing queue information with
neighboring nodes. At the same time, any MAC protocol has to allow
opportunities for all nodes to access the medium while trying to
maximize network throughput. The FD-MAC uses three mechanisms to
achieve this balance.

First mechanism is the \emph{shared random backoff}, which temporarily
couples the backoff counter for two nodes which have discovered that
they have a packet for each other. The discovery that two nodes have a
packet for each other is performed via the FD headers in every DATA
and ACK packets. If the two nodes have many packets for each other,
the FD header allows nodes to keep discovering these full-duplex
opportunities.  One possibility is that the nodes can occupy the
medium and continuously transmit to each other. However, while this
maximizes the use of full-duplex mode, it can potentially starve other
nodes. Thus, we propose that once two nodes discover that they have
more packets for each other, they use the 10-bit SRB (shared random
backoff) field in FD header to share a backoff counter with each
other.  The two nodes backoff for a common duration to stay
synchronized and at the same time allow other nodes to contend and
capture the channel. Thus the protocol balances access with the
maximal use of full-duplex mode.

The second mechanism involves nodes \emph{snooping} on headers of all
ongoing transmissions within radio range, even when the nodes have
frozen their counters during network allocation vectors (NAV).  The
packet snooping allows nodes to estimate their local topology and in
turn discover if the ongoing transmissions between the access point
and other nodes forms a clique or hidden node with themselves. If a
Mobile~$\mathsf{M}_2$ estimates that it will form a clique with the
ongoing {\sf AP} to Mobile~$\mathsf{M}_1$ flow, then $\mathsf{M}_2$
cannot exploit full-duplex since its new transmission will collide
with ongoing flow, either at {\sf AP} or at
Mobile~$\mathsf{M}_1$. However, if $\mathsf{M}_2$ -- {\sf AP} --
$\mathsf{M}_1$ forms a hidden node topology, then the asynchronous
full-duplex capabilities of FD-PHY enable injecting a new packet to
{\sf AP} while {\sf AP} is sending a packet to $\mathsf{M}_1$.

Lastly, FD-MAC uses two \emph{virtual contention resolution}
mechanisms which further balance the objective to maximally exploit
full-duplex mode to allowing access to other competing flows. The
salient mechanism is the case where the {\sf AP} looks at multiple
packets in its buffer (not just head of line packet) and statistically
decides which packet it will serve first. By looking into multiple
packets in the queue, {\sf AP} can discover more opportunities to use
the full-duplex mode. This, of course, leads to the possibility of
{\sf AP} delaying the transmission of its HOL packet which could be
problematic at higher layer protocols. So we propose to send a non-HOL
packet with vanishing probability.  Of course, this mechanism is
optional and can be completely turned off at the cost of reduced use
of full-duplex capabilities. Our experimental results show that FD-MAC
achieves a throughput gain of up to 70$\%$ over comparable half-duplex
systems. The gain is a function of distance, packet arrival pattern,
extent of contention etc.

The rest of the report is organized as follows.  In Section
\ref{sec:review}, we review the challenges and state-of-the-art in
full-duplex wireless communications. In Section~\ref{sec:phy}, we
describe the OFDM-based full-duplex physical layer (FD-PHY) study
antenna placement on actual devices and the performance of
asynchronous full-duplex. In Section~\ref{sec:mac}, we describe the
mechanisms for FD-MAC and study its behavior in prototypical
topologies, finally presenting its evaluated performance.

\vspace{-0.15cm}
\section{Review of Full-duplex Wireless} \label{sec:review}
\vspace{-0.13cm}
\subsection{Main Bottleneck in Enabling Full-duplex}

To appreciate the key challenge in achieving full-duplex wireless,
consider the two-way link shown in Figure~\ref{fig:two-way}, where the
two nodes are trying to send and receive a packet simultaneously in
the same frequency band. Node~1 has a packet for Node~2 and vice
versa. Since the situation is symmetric, we can focus our attention on
Node~1.  Assuming that the transmit and receive antenna are physically
different, the power of signal transmitted by the node by Antenna T1
causes self-interference at the receiving antenna of the node,
Antenna~R1, which can be anywhere from 15--100dB higher than the
signal of interest coming from transmit antenna Antenna T2 of
Node~2. Since most modern systems process the received signal
digitally to decode packets, the analog received signal is converted
to the digital form using an analog-to-digital converter (ADC). With
such large difference in the powers from the two signals,
self-interference and signal of interest, the finite resolution of the
ADC is the main bottleneck in enabling full-duplex communications.

\begin{figure}[h]
	\centering
        \resizebox{3.3in}{!}{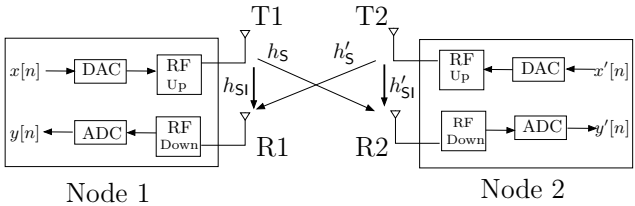}
\vspace{-0.2cm}
	\caption{A full-duplex transmission between two nodes.}
        \vspace{-0.3cm}
	\label{fig:two-way}
\end{figure}

When two radio signals impinge on the antenna, the voltage generated
at the antenna is the sum of the two signals. That voltage is
down-converted to the baseband frequency and scaled such that the sum
of the two signals occupies a voltage range (nominally denoted as
[-1,1]) such that full dynamic range of the ADC is used. This ensures
the best possible representation of the analog signal in the digital
domain. If one of the signals is much smaller than the other signal,
then it effectively gets fewer bits to represent its voltage levels
compared to the case where the smaller signal arrived at the ADC by
itself. That is because in the latter case, the automatic gain control
algorithm will scale the signal to occupy the whole ADC range and thus
allow more bits of resolution for the smaller signal by itself.

Thus, even if the SNR of the signals was {individually} high, the wide
discrepancy in their \emph{relative} amplitudes implies that the
smaller signal will have lower effective SNR in the digital domain,
leading to the stronger signal swamping the weaker signal. Thus
\emph{signal-to-interference-plus-noise ratio} (SINR) is an important
metric to determine the performance of any method for full-duplex
communications.

\subsection{Reported Methods}

To achieve full-duplex communication over reasonable distances, it is
thus important to suppress the self-interference in the analog domain
before it reaches the ADC. In 2010, two
groups~\cite{Choi:2010aa,Duarte:2010aa} reported two different
techniques to achieve approximately 60-70dB of self-interference
suppression, thereby showing the feasibility of full-duplex
transmissions. In~\cite{Choi:2010aa}, the authors proposed an antenna
cancellation method using two transmit antennas to use beamforming to
create a null at the receive antenna. Three physical antennas are
needed in the proposed method to achieve SISO full-duplex. The
prototype demonstration showed full-duplex performance for short
inter-node distances like those encountered in IEEE~802.15.4 (e.g
ZigBee) equipped devices. Perhaps a key deployment challenge is the
need for large antenna separation to achieve antenna cancellation,
especially for IEEE~802.15.4 devices which are often targeted for
small form-factor devices and thus may not have the required physical
space to accommodate the antennas.

In~\cite{Duarte:2010aa}, the authors repurposed MIMO RF chains to
generate a canceling signal and add it in analog at the antenna using
RF adder. While this technique does not have the drawback of the
additional antennas like in~\cite{Choi:2010aa}, the prototype
implementation in~\cite{Duarte:2010aa} had a very narrow bandwidth
(0.625 MHz) and thus its applicability to wide-band systems like
802.11 was not established.  




\section{Real-time Full-duplex PHY}

\label{sec:phy}
In this section, we first describe our wideband, OFDM-based,
full-duplex physical layer implemented on an off-the-shelf SDR
platform and methods to optimize antenna placement on actual
electronic devices to improve the capacity and range of full-duplex
wireless physical layer. Based on this implementation, we compare the
performance of full-duplex wireless with half-duplex physical
layers. Finally, we discuss the challenge in enabling asynchronous
full-duplex systems, and how our proposed design achieves partial
asynchronous full-duplex transmissions.

\subsection{Real-time OFDM Transceiver} \label{sec:ofdm}

The conceptual block diagram of our full-duplex physical layer is
shown in Figure \ref{fig:ofdm}. We use the narrowband technique
proposed in~\cite{Duarte:2010aa} for reducing self-interference in the
analog domain, and apply it to a wideband OFDM (orthogonal frequency
division multiplexing) system by processing each subcarier
independently.

Consider Node~1 in Figure~\ref{fig:two-way}. Denote the channel
between transmit antenna T1 and receive antenna R1 for sub-carrier $k$
as $h_k$, where $k=1,\ldots,K$ with $K$ being the total number of
sub-carriers in the OFDM system. Further, let the signal sent in
sub-carrier $k$ be denoted as $x_k$. Then the self-interference seen
at the receive antenna in the $k^{\mathrm{th}}$ subcarrier, without
any cancellation, is given by
\begin{equation}
z_{\mathsf{SI},k} = h_k*x_k .
\label{eq:before cancel}
\end{equation} 
The above representation assumes that cyclic prefix is longer than the
time delay of the multipath. This assumption is easily satisfied for
the self-interference channel since the distance between the transmit
and receive antennas of the self-interference channel is very small,
thereby resulting in very limited multipath delay. In most systems,
the cyclic prefix is designed for long distances between two nodes,
like N1 and N2 in Figure~\ref{fig:two-way}.

\begin{figure}[h]
	\centering
        \resizebox{3.2in}{!}{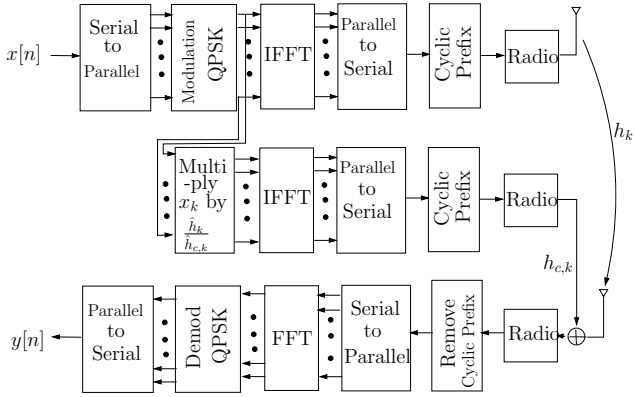}
	\caption{A block diagram of the PHY design with self
          interference cancellation.}
	\label{fig:ofdm}
\end{figure}


Following~\cite{Duarte:2010aa}, we opt for active self-interference
cancellation by using the physical layer architecture shown in Figure
\ref{fig:ofdm}, where we compute the canceling signal and cancel it
before the received signal from the receive antenna R1 reaches the
analog to digital converter. This cancellation is not performed over
the air~\cite{Choi:2010aa} but using a wired assembly and thus does
not need extra antennas. Let the wireline channel between the
cancellation transmit chain and receive antenna R1 be represented as
$h_{c,k}$ for sub-carrier $k$; note that wires are also a channels and
thus can attenuate and change phases like wireless channels.

The cancellation signal $x_{c,k}$ for the $k^\mathrm{th}$ subcarrier
is computed as 
\begin{equation}
x_{c,k} = - \frac{\hat{h}_k}{\hat{h}_{c,k}} x_k,
\label{eq:cancel signal}
\end{equation} 
where $\hat{h}_{k}$ and $\hat{h}_{c,k}$ represent the estimates of
channels $h_k$ and $h_{c,k}$. In general, the estimates have errors
and thus not equal to the quantity they are estimating.  So, the
self-interference signal received at the receive antenna \emph{after
  active analog cancellation} is
\begin{equation}
z_{\mathsf{SI},\mathsf{cancel}} = z_{\mathsf{SI,k}} + x_c.
\label{eq:after cancel}
\end{equation} 
From~(\ref{eq:after cancel}), it is clear that if the channel
estimates were perfect, the self-interference can be completely
suppressed in this technique. This is equivalent to perfect nulling in
the ideal case for the antenna cancellation technique proposed
in~\cite{Choi:2010aa}.

Since we need to estimate two sets of channels $h_k$ and $h_{c,k}$, we
can view the system as a two-transmit chain system (like in IEEE
802.11n MIMO modes) and can exploit the already available physical
layer headers in MIMO packets. Thus, no special PHY headers need to be
added to estimate the required channels to compute the canceling
signal.

We leveraged the open-source MIMO physical layer designs available at
the WARP website~\cite{warp} as the starting point for our
implementation. The open-source design occupies 10~MHz bandwidth using
64 sub-carriers and also supports 2$\times$2 MIMO transmissions.  One
of the modes in the open-source design is spatial multiplexing, where
the transmitter sends two different streams of the data to two
transmit antennas. We repurposed the spatial multiplexing mode to
implement the above scheme, where the second stream in the MIMO design
is replaced by the canceling signal $x_c$, which requires multiplying
the first signal by appropriate canceling coefficients
$\frac{\hat{h}_k}{\hat{h}_{c,k}}$. The other major component in our
design is the design of estimation procedure to obtain the required
channel estimates $\hat{h}_k$ and $\hat{h}_{c,k}$. Here again, we used
the MIMO channel estimation blocks in the open-source
design~\cite{warp} and hence the details are not provided in this
report due to lack of space. 
the

%
%
%
%

\subsection{Antenna Placement on a Mobile Devices \label{sec:antenna}}

We next investigate how full-duplex will perform on actual mobile devices. 

The form factor of the mobile device limits its antenna placement,
distance between transmit and receive antennas, and orientation of the
antennas. At present none of the small form factor mobile devices,
like smartphones, use 802.11n MIMO modes since they cannot accommodate
two RF chains on one device. Thus, we limit our attention to larger
form factor devices, like tablets and laptops.

The driving questions are how should we place the transmit and receive
antennas on a mobile device to optimize the performance of full-duplex
nodes. We consider three configurations as shown in
Figure~\ref{fig:iPad}, with each configuration including two antennas
-- one for transmit and one for receive.

\begin{figure}[ht]
\resizebox{3.5in}{!}{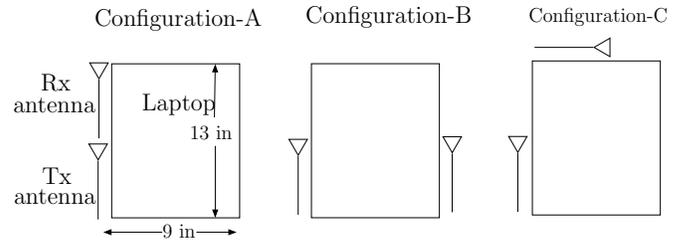}
\caption{Different antenna configurations. The same antenna
  configuration was tested in the presence and absence of the
  device}
\label{fig:iPad}
\vspace{-0.5cm}
\end{figure}

{\bf Configuration A}: While most omni-directional antennas used in
commercial devices (laptops and tablets) are reasonably
omni-directional in the far field, they are almost never truly
omni-directional in the near field. Most omnidirectional antennas have
small energy transmission along the z-axis (i.e, above and below the
antenna) ~\cite{balanis}. The antenna pattern immediately suggests a
potential deployment scenario, where the transmit and receive can be
mounted on top of each other; this is labeled as Configuration~A in
Figure~\ref{fig:iPad}.

{\bf Configuration B}: In many 802.11n equipped devices which have two
antennas to support MIMO modes, the antennas are often installed on
the opposite end of the device (like the opposite edges of the screen)
to create sufficient separation between the antennas. This is labeled
Configuration~B in Figure~\ref{fig:iPad}. The maximal separation
between the antennas creates statistically nearly-independent channels
to achieve MIMO spatial multiplexing gains. While Configuration~B was
not designed for full-duplex operation, the presence of the actual
device (e.g laptop) between the antennas has the potential to create additional
path loss between the two antennas and thereby increase the
attenuation of the self-interference.

{\bf Configuration C}: Finally, we will also test the case when one of
the antennas is installed perpendicular to the other antenna, labeled
Configuration~C in Figure~\ref{fig:iPad}. This configuration aims to
exploit the potential difference in radiation pattern along different
axes.

The experiments are performed by strapping the two 2.4 GHz 7 dBi
Desktop Omni Antenna (typical Wifi Antenna) to a iPad-sized device in different
configurations. The dimensions are shown in Figure~\ref{fig:iPad}. We
fix the transmit power at 6~dBm. For each configuration, we test the
impact of antenna configuration and the device. The results are
summarized in Table~\ref{table:configurations}. The full-duplex PHY
was implemented on WARP boards, each with three radio cards. One radio
was connected to the transmit antenna, the second was connected to the
receive antenna and the third provided the canceling signal ($x_c$)
over a wire and added in analog after the receive antenna.

\begin{table}[h]
\caption{The transmit power is 6~dBm.}
\begin{tabular}{|c|c|l|c|c|}
  \hline Config. & Device & Interference & Interference & Total 
\\  &Present & power & \small{after analog} & \small{sup-}\\ &&&
\small{cancellation}  & \small{-pression}\\ \hline 
A  & No & -28dBm & -52dBm & 58dB\\ \hline 
A  & Yes & -28dBm & -52dBm & 58dB\\ \hline 
B & No & -46dBm & -71dBm & 77dB \\ \hline 
B & Yes & -51dBm & -75dBm & 81dB \\ \hline
C & No & -40dBm & -63dBm & 69dB\\ \hline 
C & Yes & -49dBm & -73dBm & 79dB \\ \hline 

\end{tabular}
\label{table:configurations}
\end{table}

Four main results stand out from the Table~\ref{table:configurations}. 

\emph{Result 1 (Device reduces self-interference)}: Depending on the
configuration, the presence of a device (e.g laptop/iPad) can make a
significant impact on the power of self-interference, by
\emph{passively} attenuating the signal. The metallic components in a
laptop-like device can significantly attenuate the signal and thus
reduce self-interference.  In Configuration~C, device results in an
additional attenuation of 9dB attentuation compared to the case when
the device is not present. The device related attenuation is 5~dB in
Configuration~B and 0~dB in Configuration~A.

\emph{Result 2 (Best full-duplex configuration)}: The best
configuration in terms of self-interference power, with and without
analog cancellation is Configuration~B, where the self-interference
power with and without the analog cancellation is lowest compared to
other configurations. This is, in fact, great news because
Configuration~B is also the ideal configuration for MIMO
systems. Thus, there is a potential to use multiple antennas in either
MIMO or full-duplex modes in mobile devices.

\emph{Result 3 (Baseband cancelation)}:
In~\cite{Choi:2010aa,Duarte:2010aa}, baseband cancellation was also
proposed to reduce the self-interference power. In our design, we did
not implement base-band cancellation due to lack of sufficient FPGA
logic on our WARP boards, but we can still achieve a self-interference
suppression which is \emph{more} than the prior work due to added
suppression by the device.



\emph{Result 4 (RF requirements for canceling signal path)}: The
self-interference power before analog cancellation in all
configurations is more than 30dB. For example, in Configuration~A, the
received power with device is -28dBm for transmit power of 6dBm, which
implies 34dB loss in signal power when the self-interference reaches
receive antenna. This implies that the canceling transmit RF chain
does \emph{not} require a power amplifier, because the canceling
signal travels over a wire and thus suffers only minor attenuation. In
fact, we had to install 40~dB attenuators on our off-the-shelf radio
cards, which essentially removed all the power amplification by the
power amplifiers. This is again an encouraging news, which shows that
the full-duplex transceiver needs one full transmit chain
(up-converter for transmit antenna), one radio chain (down-converter
for receive antenna) and a partial transmit chain without power amplifier (for canceling
signal). Thus, compared to SISO transceiver (one transmit and one
receive RF chain), full-duplex only needs the additional partial
transmit chain.

%

\subsection{Asynchronous Full-duplex \label{sec:channel estimation}}

So far, the PHY analysis in prior
works~\cite{Choi:2010aa,Duarte:2010aa} and in
Section~\ref{sec:antenna} has been motivated by two nodes exchanging
packets with each other as shown in Figure~\ref{fig:two-way}. However,
full-duplex can be employed in more general cases. Consider the hidden
node topology in Figure~\ref{fig:FD-capable}(b), where $\mathsf{M}_2$
is out of radio range of $\mathsf{M}_1$. Assume {\sf AP} has a packet
for $\mathsf{M}_1$ and $\mathsf{M}_2$ has a packet for {\sf AP}. In
this case, since the {\sf AP} has to be a full-duplex node, the key
question is if the full-duplex mode can be enabled in an
\emph{asynchronous} manner. That is, can a new flow be added once a
flow starts transmission. In the hidden node example, there are two
possibilities for {\sf AP}: (a)~start receiving a packet from
$\mathsf{M}_2$ after having initiated a transmission to
$\mathsf{M}_1$, (b)~start a transmission to $\mathsf{M}_1$ while
receiving a packet from $\mathsf{M}_2$.

\noindent
\emph{\bf A new reception while transmitting}: Assume that {\sf AP} is
actively transmitting to $\mathsf{M}_1$ and is continuously operating
its analog canceler to suppress its own self interference. This
ensures that when $\mathsf{M}_2$ starts a packet, it can be decoded by
{\sf AP}'s receiver. The key challenge is that {\sf AP} has to
estimate the channel between $\mathsf{M}_2$ and {\sf AP} in the
presence of self-interference, which is required to be able to decode
$\mathsf{M}_2$'s packet at {\sf AP}. In almost all current systems,
even with multiple users, this training is performed without any
(intentional) interference.

However, to enable asynchronous full-duplex, we are required to
estimate the channel between $\mathsf{M}_2$ and {\sf AP} in the
presence of self-interference caused by {\sf AP}'s ongoing
transmission. We label the physical layer channel estimation in the
presence of ongoing transmission as \emph{dirty estimation}, and
quantify the loss compared to the conventional systems, all of which have \emph{clean estimation}.


\begin{table}[h]
\begin{center}
\caption{Each packet has a payload of 324 bytes and was QPSK-encoded.
  Signal transmit power was fixed at 6dBm. A total of $1.3\times10^6$ bits were transmitted.  \label{table:clean_vs_dirty}}
\begin{tabular}{|c|c|c|c|} \hline
SINR  & BER & BER \\ & dirty & clean \\ {\small (with canceler)} &
estimation & estimation \\ \hline
 18~dB & $2\times 10^{-6}$ & $0$ \\ \hline
 14~dB &  $4 \times 10^{-4}$   & $ 1.4 \times 10^{-4}$ \\ \hline
 11~dB  &  $9 \times 10^{-3}$   & $ 1.8 \times 10^{-3}$ \\ \hline
8~dB &  $2.4 \times 10^{-2}$ & $5 \times 10^{-3}$ \\ \hline
7~dB & $2.5 \times 10^{-2}$ & $9 \times 10^{-3} $ \\ \hline
\end{tabular}
\end{center}
\end{table}

In Table~\ref{table:clean_vs_dirty}, we report the results for
different values of SINR which were achieved by changing the distance
between the two nodes $\mathsf{M}_2$ and {\sf AP}. From
Table~\ref{table:clean_vs_dirty}, it is clear that estimating the
$\mathsf{M}_2 \rightarrow $ {\sf AP} channel in the presence of self-interference
increases the bit error rate (BER) for all distances. The impact is
worse as the SINR reduces; for high SINR, there is hardly any
measurable loss and for low SINR, the BER in dirty estimation system
can be 6 times compared to clean estimation, which turns out to be up to 3~dB loss in effective SINR for the asynchronous packet. \emph{This implies that
  the capacity of the full-duplex transmission is reduced if
  full-duplex is used in this asynchronous mode.} \\

\noindent
\emph{\bf A new transmission while receiving}: Now we consider the converse case, where {\sf AP} is already receiving a packet from M$_2$ and intends to send a packet to M$_1$ to leverage its full-duplex capabilities. \emph{Unfortunately, this mode cannot be enabled reliably.}

The key challenge is calculation of the self-canceling signal in the
presence of an ongoing reception. To compute the canceling signal
$x_c$, we need to estimate the channel coefficients $h_k$ and
$h_{c,k}$. If the MIMO PHY header is transmitted (as described in
Section~\ref{sec:ofdm}) while PHY is receiving a packet, then the
large uncanceled self-interference will completely swamp the ongoing
reception. This is because self-interference \emph{before} cancelation
is almost always much bigger than signal of interest (as also discussed in Section~\ref{sec:review}). While receiving
the packet, the automatic gain control (AGC) is set to ensure that the
incoming signal occupies the full dynamic range of the
analog-to-digital converter (ADC). Thus the process of estimating the channels to establish
canceling signal causes a ``self-collision" at the receiver.

A possible approach is to backoff on how much of the dynamic range is
occupied by the receiving packet, thus allowing a big uncanceled
signal to not completely destroy the packet. The drawback of lost bits
of resolution is that the quantization noise of the receiver is
increased, which decreases its effective SINR, increasing BER and
thereby reducing overall throughput.

Another approach will be not estimate the self-interference channel
and simply use older estimates for the desired channels. In our
experiments, the self-interference channel with a device in the middle
had sufficient variations over time, which implies that
self-interference canceler can end up doing more harm than good if it
has outdated channel estimates. This again, leads to the same
situation where full-duplex cannot be enabled reliably.

\emph{Result 5 (Allowable asynchronous modes)}: The key result is that asynchronous full-duplex can be enabled to receive while transmitting (with some loss in the performance of receiving packet) but not transmit while receiving.

\section{MAC protocol design}  \label{sec:mac}

In this section, we will describe Full-Duplex Medium Access Protocol
(FD-MAC) which uses the full-duplex-capable physical layer described
in Section~\ref{sec:phy}. We will limit our attention to
infrastructure based systems and focus on the scenario involving one
access point ({\sf AP}). This will allow us to define the fundamental
elements of a full-duplex MAC protocol.

\subsection{Challenges in MAC Design}
\label{subsec:challenge_in_mac}

The first challenge in designing full-duplex MAC is identification of
the nodes which can engage in a full-duplex mode. In any network of
multiple nodes, multiple flows with random arrivals exist at the same
time, leading to random instances when full-duplex can be used.

The second challenge is imposed by the physical layer. From
Section~\ref{sec:channel estimation}, either full-duplex has to be
performed synchronously between two nodes (a packet exchange) or can
be done asynchronously only if a full-duplex node receives a packet
while transmitting a packet to another node.  Any MAC design has to
respect this constraint in its design.

The third challenge is shared by any MAC protocol (full or
half-duplex) and is to provide opportunity to all nodes to access the
medium. Thus, the access protocol should not unduely favor full-duplex
opportunities over half-duplex flows.


\subsection{Overview of FD-MAC}
\label{subsec:mac_overview}

In the infrastructure-based network, all flows have either {\sf AP} as
their source or destination. Thus, at any given time, a maximum of two
flows can be active among full-duplex capable nodes. The two possible
scenarios which leverage full-duplex capabilities are shown in
Figure~\ref{fig:two-node} and ~\ref{fig:hiddenflows1}, where (i)~{\sf
  AP} and mobile node $\mathsf{M}_1$ are exchanging packets or
(ii)~{\sf AP} is sending and receiving a packet simultaneously from
two mobile nodes $\mathsf{M}_1$ and $\mathsf{M}_2$, which are hidden
from each other.

\begin{figure}[h]

\subfigure[{The simplest network with 2
    nodes.}]{\label{fig:two-node}\resizebox{0.9in}{!}{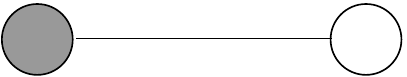}}
\hspace{1cm}
\subfigure[{Both the mobile nodes are connected to the $\mathsf{AP}$
    but are not in the radio range of one
    another}]{\label{fig:hiddenflows1}\resizebox{1.5in}{!}{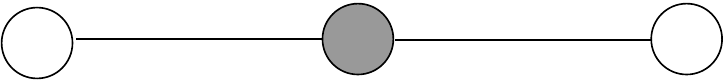}}

\subfigure[{All three nodes are in radio range of each
    other}]{\label{fig:clique}\resizebox{1in}{!}{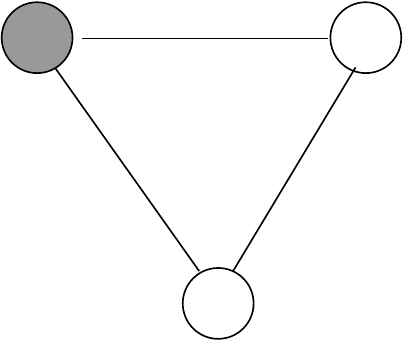}}\hspace{1cm}
\subfigure[{$\mathsf{M}_2$ and $\mathsf{M}_3$ are hidden to
    $\mathsf{M}_1$}]
          {\label{fig:threenode}\resizebox{1.5in}{!}{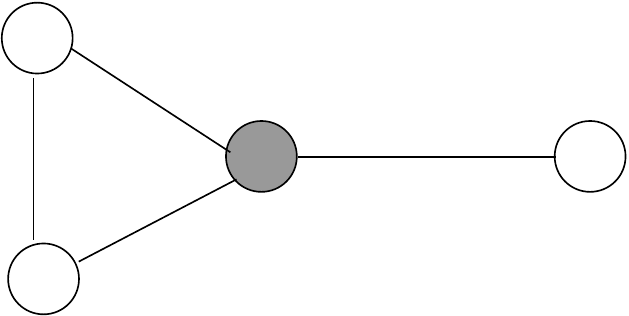}}
\caption{ A line connecting any two nodes indicates
  that they are in radio range of one another}
\label{fig:FD-capable}
\end{figure}

 FD-MAC is a random access protocol, which will use most of the
 dominant elements of the IEEE~802.11 DCF. However, while IEEE~802.11 is
 CSMA/CA, collision avoidance in FD-MAC is done selectively to
 leverage full-duplex opportunities. FD-MAC introduces following three
 new protocol elements.
 
 \emph{Shared random backoff}: When two nodes, say {\sf AP} and
 $\mathsf{M}_1$ in Figure~\ref{fig:two-node}, are in a situation where
 they have many packets for each other and thus truly exploit
 full-duplex, they do not continuously capture the medium in order to
 allow other nodes to send or receive from {\sf AP}. Instead they
 agree on a shared random back-off which allows other nodes to contend
 for the medium. If no one else wins, the two nodes can continue
 with their full-duplex transmission.
 
 \emph{Snooping to discover full-duplex opportunities}: In FD-MAC,
 nodes decode headers of all ongoing transmissions, even when network
 allocation vector NAV is non-zero. This allows the nodes to estimate
 the local topology and initiate full-duplex opportunistically.

 \emph{Virtual contention resolution}: FD-MAC also has two virtual
 contention mechanisms to balance use of the
 full-duplex mode with access for all nodes in the network. 
 
 While the FD-MAC can be used with or without RTS/CTS, we will only
 describe for the more popular use case of infrastructure mode of
 802.11 which does not use RTS/CTS.

\subsection{FD-MAC Packet structure}
\label{subsec:mac_pkt_struct}

We adopt IEEE~802.11 packet structure and add a new FD header, for
managing full-duplex transmissions as shown in
Figure~\ref{fig:pkt_struct}.
%
%
Each packet contains a PHY header, a MAC header, a full-duplex header,
a payload and a cyclic redundancy check (CRC). Except for the
full-duplex (FD) header, all other fields are identical to IEEE~802.11
packets. We briefly explain the fields which are essential to describe
FD-MAC.

The PHY header has a preamble and the training symbols necessary for
the functioning of the physical layer. The existing elements of the
MAC header that we use in our FD-MAC protocol are Duration ID denoting
the duration (DUR) of the packet, source address (SA), destination
address (DA) and FRAG (denoting if there are more fragments of the
same packet in line for the destination). The MAC header distinguishes
between data packet and acknowledgement. For simplicity of description
of the FD-MAC protocol, data packets will be referred as DATA and
acknowledgement as ACK.

The FD header has a one-bit field to distinguish packet type (DUPMODE)
which can either assume values HD (indicating that it is a half-duplex
packet) or FD (indicating that it is a full-duplex packet). Then there
is a one-bit field, Head-of-line (HOL), indicating that the next
packet in the buffer is for the destination of the current packet.
The current 802.11 MAC header has a field labeled `more data' in Frame
Control Field of MAC header, but to avoid any conflict with other uses
this field, we have defined HOL in the FD header. The overall overhead
increase is minimal since the HOL is only 1-bit long.

The next field reveals the duration of head of line packet, DURNXT,
and is useful when HOL = 1. It 2 bytes long. The next field is meant
for revealing duration of the full-duplex exchange, DURFD. It too is 2
bytes long.


The next one-bit is a Clear-To-Send (CTS) indicating that destination
of the current packet can send a packet to source of the current
packet. Finally the FD header has a field for a 10-bit number which is
the Shared Random Backoff (SRB).

Fields DURNXT and DURFD are needed in order to counter the hidden node
problem in infrastructure mode of 802.11. They are \emph{optional} and
in their absence, the FD header is only 13 bits.

\begin{figure}[h]
\centering
\resizebox{2.2in}{!}{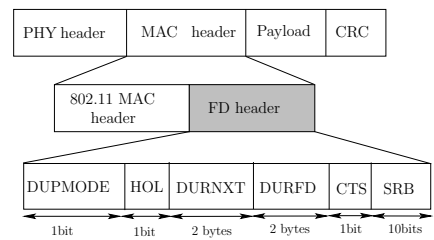}
\caption{Structure of the packet being used for the FD-MAC
  protocol \label{fig:pkt_struct}}
\end{figure}

\newpage
\subsection{Shared Random Backoff \label{sec:srb}}

Consider the most basic two-node example shown in
Figure~\ref{fig:two-node}. It is possible that at any given time
either both nodes have a packet for each other or only one node has a
packet for the other. Note that in this case, asynchronous full-duplex
is not possible because of the PHY constraints
(Section~\ref{sec:channel estimation}), where a node cannot start a
new transmission while it is receiving a packet. Thus, nodes have to
find a way to synchronizing their transmissions, such that they can
estimate the channel coefficients for maximal self-interference
cancellation (as discussed in Section~\ref{sec:ofdm}).

To maximize the use of full-duplex mode while respecting the
constraints imposed by the physical layer, FD-MAC proceeds as
follows. Assume that the nodes contend for the medium since they do
not know if both nodes have a packet for each other or not. Without
loss of generality, assume that {\sf AP} wins the contention
resolution. Then if the {\sf AP} has another packet lined up in the
buffer for $\mathsf{M}_1$, it sets HOL=1 in the DATA packet . Here
$\mathrm{SRB}_{\mathsf{AP}} = 0 $ and the DUPMODE = HD. Thus, the
first packet in a two-way exchange is half-duplex.


\begin{figure*}[h]
\resizebox{6.0in}{!}{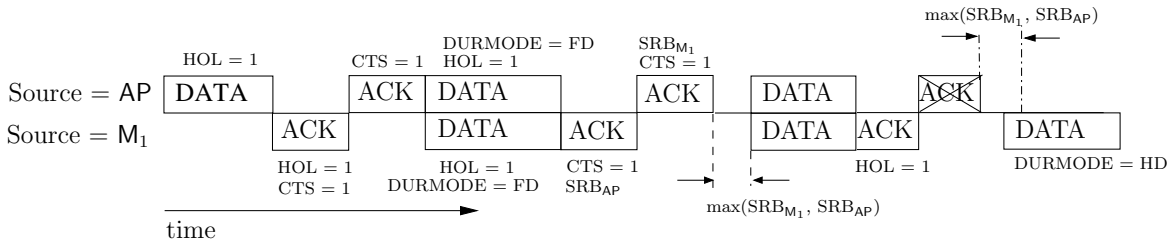}
\vspace{-0.1cm}
\caption{Timeline of packets sent from $\mathsf{AP} \rightarrow
  \mathsf{M}_1$ and $\mathsf{M}_1 \rightarrow \mathsf{AP}$. The
  relevant fields for decision making are listed above and below the
  packets.} 
\label{fig:two-waysetup}
\vspace{-0.4cm}
\end{figure*}

If $\mathsf{M}_1$ receives the DATA successfully and has a packet for
{\sf AP}, it sends and ACK packet with HOL=1 and DURNXT set to the
length of the head of the buffer packet. Also, SRB=0, CTS = 1. After
receiving the ACK, both nodes know that they can initiate a
full-duplex. The PHY needs $\mathsf{AP}$ to train its
self-interference channel, and thus $\mathsf{AP}$ sends an ACK packet,
with HOL = 1, and also reveals DURNXT, and set SRB=0, CTS=1. Now the
two nodes are set to be in full-duplex. They wait for
$\max(\mathrm{SRB}_{\mathrm{DATA}},\mathrm{SRB}_{\mathrm{ACK}})$
(which is = 0 at this stage) and then send their respective DATA
packets with the DUPMODE = FD, and\\ DURFD=
$\max(\mathrm{DURNXT}_{\mathsf{AP}},
\mathrm{DURNXT}_{\mathsf{M}_1})$. Each node sends an ACK only at the
end of DURFD duration. Also, {\sf AP} always sends the ACK after the
$\mathsf{M}_1$ in full-duplex mode, which allows hidden nodes to
contend in the medium at the end of the ACK from $\mathsf{AP}$; see
Section~\ref{sec:snoop}.

After one full-duplex transmission, it is possible that both nodes
still have more packets for each other, which they will discover by
setting the FD header fields as described above. However, if the two
nodes continue to occupy the medium without any breaks, then other
nodes will get completely starved. On the other hand, if the two
nodes know they have a packet for each other but give up the medium
for other nodes, they will have to go through a contention resolution
again followed by one half-duplex packet. Thus, it is important for
nodes to retain the knowledge of \emph{queue state} which they obtain
by above hand-shaking enabled by FD header.

So, we introduce the idea of \emph{shared random backoff}~(SRB), where
{\sf AP} and $\mathsf{M}_1$ handshake on the random delay they will both wait
before resuming full-duplex mode. In the ACK sent after receiving
first full-duplex packet, {\sf AP} picks a random backoff from
$[0,\mathsf{CW}_{\max,\mathsf{AP}}]$ where
$\mathsf{CW}_{\max,\mathsf{AP}}$ is the current maximum contention
window width for {\sf AP} and places that number in SRB. The mobile
node $\mathsf{M}_1$ also picks a random backoff from its own maximum
contention window $[0,\mathsf{CW}_{\max,\mathsf{M}_1}]$, and places it
the SRB field of its ACK packet FD header.

After the two nodes have finished sending ACKs, they wait for
$\max(\mathrm{SRB}_{\mathsf{AP}},\mathrm{SRB}_{\mathrm{M}_1})$.
In the 802.11 DCF, backoff countdowns are paused by carrier-sense
events. In our work, we require distributed nodes to independently
count down for the same duration and, as such, cannot employ this
pausing mechanism since they each might see independent channel
busyness events. Hence, we propose a different kind of behavior for
the shared random backoff; nodes do not pause their backoff countdowns
in the presence of energy on the medium but instead perform one final
idle-for-DIFS check at the end of the interval ensure that there is
nothing currently using the medium when they are about to transmit. If
no other node in the network wins the medium before this shared
backoff counter expires, the two nodes enter the full-duplex mode
again and continue the above process till they have packets for each
other. A timeline of the events is shown in Figure 6. Note that the
protocol requires AP and $\mathsf{M}_1$ to wait for at least
$\max(\mathrm{SRB}_\mathsf{AP}, \mathrm{SRB}_{\mathsf{M}_1})$ before
transmitting, however it can tolerate more delay in start of DATA
packets of $\mathsf{AP}$ as the PHY layer as already estimated the
required channels. However if another node wins the medium before the
expiry of the calculated backoff, then both $\mathsf{AP}$ and
$\mathsf{M}_1$ purge their knowledge about the other nodes and start
completely afresh.

The reason to purge the states is because upon discovering full-duplex
opportunities with another node, say $\mathsf{M}_2$, the $\mathsf{AP}$
will modify the ordering of packets in its buffer to place packets
destined for $\mathsf{M}_2$ in front of the buffer. In Section 4.6, we
discuss the idea of reordering the buffer in more detail. Another
reason for this purge is to account for the previously discussed
modification to the backoff process. In the presence of other traffic,
the shared backoff will effectively be cancelled despite the removal
of the explicit pausing mechanism. Thus, the only difference between
traditional backoffs and our shared backoffs is the fact that our
full-duplex nodes will not pause their backoffs in the presence of
undecodable energy on the medium.  at AP in more detail.


\emph{Failure of} DATA \emph{or} ACK: In a full-duplex exchange if a
node does not decode DATA correctly, it does not send the
corresponding ACK. At this point, synchronization of backoffs is not
possible and since both nodes have not received at least one of DATA
or ACK, both $\mathsf{AP}$ and $\mathsf{M}_1$ purge the information
about queue state of the node and contend for the medium once they do
not detect any energy on the medium. On the other hand if only one of
the ACK fails, the node not receiving the ACK purges its knowledge of
the queue and contends for the medium at the end of two ACK periods
after the DATA packet exchange finishes. It is then a case of physical
medium contention by the nodes with one of nodes having a the backoffs
fixed to $\max(\mathrm{SRB}_{\mathsf{AP}},
\mathrm{SRB}_{\mathsf{M}_1})$ and others having a random backoff. Both
ACK failure simply calls for purging queue state information and thus
results in another 802.11 -like contention. Therefore in the poor
channel conditions case too, the FD-MAC has a throughput at least as
much as that of 802.11 (minus the throughput loss due to additional FD
header). \vspace{0.3cm}



\subsection{Snooping to Leverage FD Mode \label{sec:snoop}} 

Consider the case of three nodes, one {\sf AP} and two mobile nodes
$\mathsf{M}_1$ and $\mathsf{M}_2$. With three nodes such that both
mobile units can communicate with the {\sf AP}, there are two possible
topologies: (i)~all nodes can hear each other and thus forming a
clique and (ii)~$\mathsf{M}_1$ and $ \mathsf{M}_2$ are not in the
radio range of each other and thus hidden from each other. We discuss
how snooping headers of the ongoing transmissions can help nodes
identify opportunities to leverage full-duplex modality.

\begin{figure*}
\resizebox{5.5in}{!}{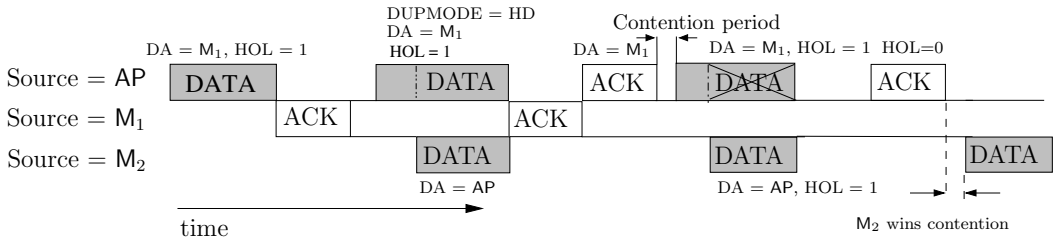}
\vspace{-0.1cm}
\caption{ $\mathsf{AP} \rightarrow \mathsf{M}_1$ and $\mathsf{M}_2$ is
  hidden from $\mathsf{M}_1$ . ACKs from $\mathsf{M}_1 \rightarrow
  \mathsf{AP}$ are not received at $\mathsf{M}_2$.  The dashed lines
  in DATA packet of $\mathsf{AP}$ signify the end of the header which
  $\mathsf{M}_2$ can decode. Corruption of DATA implies no ACK from
  receiver}
\label{fig:hiddensetup}
\vspace{-0.4cm}
\end{figure*}

We note that there is no explicit topology discovery mechanism in
FD-MAC. Thus, nodes estimate the topology by overhearing packets as
follows. Assume {\sf AP} sends a DATA packet to $\mathsf{M}_1$. Since
$\mathsf{M}_2$ is associated with {\sf AP}, so it can decode the
headers and knows that the packet is addressed to M$_1$. If the ACK
from $\mathsf{M}_1$ is overheard by $\mathsf{M}_2$, then
$\mathsf{M}_2$ concludes that it forms a clique topology with
$\mathsf{M}_1$. Else it concludes that it is in hidden-node topology
with $\mathsf{M}_1$. Note that $\mathsf{M}_2$ can make an error in its
estimation due to random channel induced errors causing either the
DATA or ACK to drop, each leading to a wrong conclusion at
$\mathsf{M}_2$. However, since MAC headers can be encoded at
base-rate, the probability of making errors is often negligibly small.

If $\{\textrm{M}_1, \textrm{M}_2, \mathsf{AP} \}$ form a clique, then
the only possible full-duplex combinations are $\mathsf{AP}
\rightleftarrows \textrm{M}_1$ and $\mathsf{AP} \rightleftarrows
\textrm{M}_2$. The combinations $\{ \mathsf{AP} \rightarrow
\textrm{M}_1, \textrm{M}_2 \rightarrow \mathsf{AP}\}$ and $\{
\mathsf{AP} \rightarrow \textrm{M}_2, \textrm{M}_1 \rightarrow
\mathsf{AP}\}$ are not possible because they cause collisions (two
simultaneous incoming packets) at one of the mobile nodes due to the
topology being a clique.

Now consider the case of hidden node topology. In this case, all four
full-duplex combinations are possible: (i)~$\mathsf{AP}
\rightleftarrows \textrm{M}_1$, (ii)~$\mathsf{AP} \rightleftarrows
\textrm{M}_2$, (iii)~$\{ \mathsf{AP} \rightarrow \textrm{M}_1,
\textrm{M}_2 \rightarrow \mathsf{AP}\}$ and (iv)~$\{ \mathsf{AP}
\rightarrow \textrm{M}_2, \textrm{M}_1 \rightarrow \mathsf{AP}\}$. We
have discussed how to establish the first two full-duplex
opportunities, (i) and (ii), in Section~\ref{sec:srb}. The third and
fourth cases are mirror reflections of each other, so we can focus on
any one of the two. Without loss of generality, consider case (iii).

We first recall that there is a PHY-imposed constraint that a node
cannot initiate a new transmission if it is already receiving a packet
from another node. Thus, since only {\sf AP} will be in full-duplex
mode in case~(iii), this case is only possible if {\sf AP} begins to
send its packet to $\mathsf{M}_1$ first. Assume that that is the case
where {\sf AP} wins the contention resolution and begins sending its
packet to $\mathsf{M}_1$. By snooping on the HOL field of the FD
header, $\mathsf{M}_2$
can learn if {\sf AP} has another packet for $\mathsf{M}_1$ or not. If
{\sf AP} does have a HOL line packet for $\mathsf{M}_1$, then
$\mathsf{M}_2$ can tranmit a packet to $\mathsf{AP}$ while
$\mathsf{AP}$ transmit its next packet to $\mathsf{M}_1$, if (a)
$\mathsf{M}_1$ is not the radio range of $\mathsf{M}_2$ and (b)
$\mathsf{M}_1$ should not be attempting to achieve $\mathsf{AP}
\rightleftarrows \mathsf{M}_1$. 

In order to ensure (a) $\mathsf{M}_2$ waits for one ACK duration after
the finish of DATA packet from {\sf AP}. If $\mathsf{M}_2$ does not
receive the ACK, it assumes that $\mathsf{M}_1$ is not its radio
range. 


In order to ensure (b), $\mathsf{M}_2$ does not contend for the medium
and allows the $\mathsf{AP}$ to capture the media. It then decodes the
FD header of DATA packet being sent from $\mathsf{AP}$. If its
destination is $\mathsf{M}_1$ and the DUPMODE is HD, then
$\mathsf{M}_1$ can transmit its own packet to $\mathsf{AP}$. It
decodes the duration DUR of $\mathsf{AP}$'s packet and fragments its
packet to ensure it ends no later than $\mathsf{AP}$'s
transmission. It also sets the FRAG =1 in its packet. The
fragmentation is necessary to avoid collisions with the ACK from
$\mathsf{M}_1$. The ACK from $\mathsf{AP}$ will arrive one ACK period
after the finish of the DATA packet. The same procedure continues as
long as $\mathsf{AP}$ has a packet for $\mathsf{M}_1$, $\mathsf{M}_2$
has a packet for $\mathsf{AP}$, and $\mathsf{M}_1$ does not have a
packet for $\mathsf{AP}$. Event timeline is shown in the Figure
\ref{fig:hiddensetup}.

At any point of time if $\mathsf{M}_1$ has a packet for $\mathsf{AP}$,
it will coordinate with $\mathsf{AP}$ via ACKs to enable $\mathsf{AP}
\leftrightarrows \mathsf{M}_1$, and $\mathsf{M}_2$ can discover this
setup if the DUPMODE=FD for the DATA packet from $\mathsf{AP}$ to
$\mathsf{M}_1$. Moreover, DURFD will let $\mathsf{M}_2$ know that it
should not contend for the medium at least for DURFD + 2ACK
periods. This gets rid of unnecessary collisions if packets of
$\mathsf{AP}$ are much smaller than that of $\mathsf{M}_1$.



\subsection{Virtual Contention Resolution \label{sec:virtual}}

In the previous two sections, we introduced methods to allow mobile to
{\sf AP} flows to get a chance to contend (Section~\ref{sec:srb}) and
discover opportunities to exploit full-duplex capabilities at PHY
(Section~\ref{sec:snoop}). In this section, we introduce two more
mechanisms which allow (i)~{\sf AP} to break away from a full-duplex
handshake to send packets to other nodes, since {\sf AP} can have
downlink flows for any mobile node associated with it, and (ii)~reduce
the probability of collisions in snooping based full-duplex access.

First consider the case where {\sf AP} in a full-duplex packet
exchange with a mobile node $\mathsf{M}_1$.  In standard 802.11 MAC
protocol, always the packet at the front of the buffer is transmitted,
i.e. the depth of the MAC buffer is one.  In order to further increase
possibility of operating in full-duplex, the $\mathsf{AP}$ can have a
larger MAC buffer such that it has more chances to find a packet for
$\mathsf{M}_1$ as long as the node $\mathsf{M}_1$ has a packet to send
to the $\mathsf{AP}$. It does so by placing the next available packet
destined for $\mathsf{M}_1$ in front of its buffer, as shown in
Figure~\ref{fig:buffer}, making MAC no longer a FIFO layer.
$\mathsf{Bufdepth}$ is a parameter that can be increased to achieve
full-duplex exchange. We note that if FIFO operation is desired then
$\mathsf{Bufdepth}$ can be chosen to be one and hence this mode is
\emph{optional}.

Increasing the depth of the buffer improves the chance of operation in
full-duplex. On the flip side, it can starve transmission of packets
to other mobile nodes. In order to break away from the full-duplex
handshake and allow $\mathsf{AP}$ to send packets, virtual contention
is arranged between the destination of the current head of the buffer
and the destination with whom {\sf AP} engaged in full-duplex
exchange. Upon discovering an opportunity of a full-duplex exchange
with $\mathsf{M}_1$, $\mathsf{AP}$ searches for a packet with
$\mathsf{M}_1$ as its destination in its buffer, and sends it if
found. After the first full-duplex exchange, {\sf AP} searches through
2nd to $\mathsf{Bufdepth}$ packets in the buffer and with a
probability $p_{\mathrm{pick}}$ picks the first packet with
destination=$\mathsf{M}_1$ as the new head of line packet. Since the
probability of picking $k$ consecutive out-of-order packets decays
geometrically as $p_{\mathrm{pick}}^k$, the $\mathsf{AP}$ chooses to \emph{not}
send head-of-line packets with a fast decaying probability.

\begin{figure}[h]
  \resizebox{2.9in}{!}{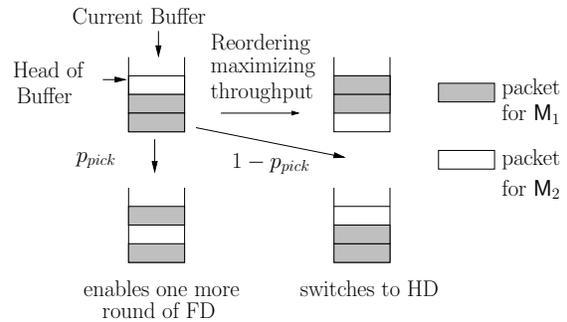}
\caption{Virtual contention resolution between packets in the buffer
  of $\mathsf{AP}$ with $\mathsf{Bufdepth} = 3$. 
  Virtual contention is a probabilistic reordering of the MAC buffer at the end of every
  full-duplex exchange}
\label{fig:buffer} \vspace{-0.4cm}

\end{figure}


Second, consider the case where multiple nodes are snooping on ongoing
transmissions by {\sf AP} as described in Section~\ref{sec:snoop}, say
$\mathsf{M}_3$ in addition to $\mathsf{M}_2$ as shown in
Figure~\ref{fig:FD-capable}(d). If both $\mathsf{M}_2$ and
$\mathsf{M}_3$ are hidden from $\mathsf{M}_1$, then they will both
send a packet to {\sf AP} at the same time and end up colliding at
{\sf AP} since {\sf AP} can only receive one packet at a time. Thus,
it is important that there is a mechanism to avoid such
collisions. Since $\mathsf{M}_2$ and $\mathsf{M}_3$ do not know how
many nodes are there which may try to contend, they only send a packet
to use the full-duplex mode at {\sf AP} probabilistically. That is,
each node which detects a full-duplex opportunity, sends the packet
with probability $p_i$, where $p_i$ is computed based on the current
maximum backoff window as $p_i = \frac{\beta}{\mathsf{CW}_{\max}}$,
where $\beta$ is a pre-chosen constant which controls the
aggressiveness in the system.

The motivation for using $p_i \propto \frac{1}{\mathsf{CW}_{\max}}$ is
that each node can use their current maximum contention window as a
proxy for amount of expected competition in the system. Since each
node's neighborhood is different, all nodes face a different amount of
contention on the average. Of course, it is possible to fix $p_i = p$
where $p$ is pre-chosen and allows equal chance for each nodes. 

 \emph{ Result 6 (Impact of larger buffer depth)}: Increasing the
 buffer depth at $\mathsf{AP}$ increases the throughput. The increase
 in throughput comes at the cost of increased delay due to packet
 reordering. In order to understand the tradeoff between delay and
 throughput due to larger than one $\mathsf{Bufdepth}$, and the
 probability $p_{\mathrm{pick}}$ we simulate the buffer of the
 $\mathsf{AP}$ with packets for 5 mobile nodes. All nodes always have
 packets for the $\mathsf{AP}$, and in radio range of one
 another. Thus full-duplex exchange is always possible and can be
 broken only via virtual contention. Only type of contention allowed
 was virtual contention in the buffer of $\mathsf{AP}$. For every
 $\mathsf{Bufdepth}$, $p_\mathrm{pick}$ was ranged from 0 to 1. The
 $\mathsf{AP}$ had uniform traffic for all nodes with packets lined up
 in an arbitrary order. Figure~\ref{fig:ThptDelay} shows a plot of
 throughput vs. average delay (for the head of the buffer packet) for
 different buffer depths. A key finding that the throughput and
 average delay are linearly related. Larger $\mathsf{Bufdepth}$ can
 help in improving the throughput at the cost of delay. Also, it is
 often possible to obtain the same (throughput, average delay) pair
 for smaller $\mathsf{Bufdepth}$ by simply increasing the probabilty
 of reordering, $p_{\mathrm{pick}}$.

\begin{figure}[ht]
\vspace{-2.9cm}\centering
\includegraphics[scale=0.4]{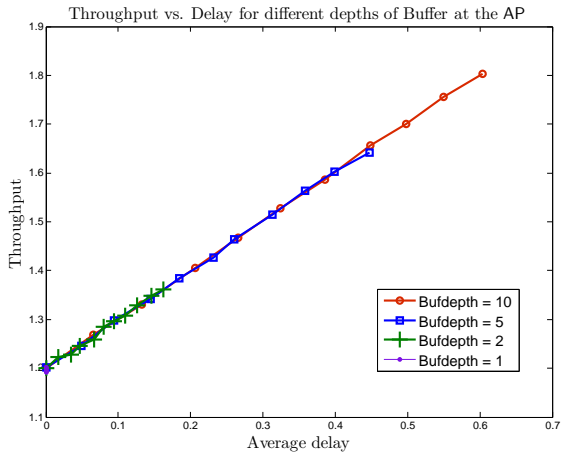}
\label{fig:ThptDelay}
\vspace{-3.1cm}
\caption{The throughput is normalized, with a maximum and minimum
  being 2 (all FD packets) and 1 (all HD packets).$\mathsf{Bufdepth} =
  1$ implies 0 delay}
\vspace{0cm}
\end{figure}

The protocol description is now complete. In the next section, we
consider an example topology to understand how all the proposed
mechanisms in FD-MAC come in play.

\subsection{State Transitions in FD-MAC \label{sec:state transitions}}

Consider the clique topology shown in Figure~\ref{fig:clique}. Assume
that there are four flows in the network, $\mathsf{AP} \rightarrow
\mathsf{M}_1$, $\mathsf{AP} \rightarrow \mathsf{M}_2$, $\mathsf{M}_1
\rightarrow \mathsf{AP}$, $\mathsf{M}_2 \rightarrow \mathsf{AP}$. Four
flows bring forth the possibility of two full-duplex scenarios
$\mathsf{AP} \leftrightarrows \mathsf{M}_1$ and $\mathsf{AP}
\leftrightarrows \mathsf{M}_2$.



Figure~\ref{fig:states} shows the mechanisms which allow the network
to go from one mode to another. Each transition is enabled by the
features introduced by the FD-MAC protocol. The three node network
with clique topology can transition from one full-duplex mode
i.e. $\mathsf{AP}\leftrightarrows \mathsf{M}_1$ to $\mathsf{AP}
\leftrightarrows \mathsf{M}_2$ only through half-duplex modes. This is
so because the first packet in two-way full-duplex exchange, as
discussed in Section~\ref{sec:srb}, is always a half-duplex
packet. Suppose that network is in the mode $\mathsf{AP}
\leftrightarrows \mathsf{M}_1$. From this full-duplex mode, the
network can transition to a half-duplex mode due to different reasons:
(a) if at least one of $\mathsf{AP}$ or $\mathsf{M}_1$ has no more
packets in the buffer for the other, i.e. if(HOL$_\mathsf{AP} = 0$ or
HOL$_{\mathsf{M}_1} = 0$), both the $\mathsf{AP}$ and $\mathsf{M}_1$
naturally give up full-duplex mode (b) any of the DATA or ACK packets
is not decoded right - failure in reception leads to purging of queue
states of other nodes to start 802.11 type contention (c) the packets
with $\mathsf{M}_2$ as destination win the virtual contention
resolution allows the network to break away from full-duplex entering
a $\mathsf{AP} \rightarrow \mathsf{M}_2$ mode, and (d) $\mathsf{M}_2$
wins the physical contention during the silent shared random backoff
period, thus initiating $\mathsf{M}_2 \rightarrow \mathsf{AP}$.

All the half-duplex modes can switch among each other with the 802.11
protocol. Consider the half-duplex mode $\mathsf{M}_1 \rightarrow
\mathsf{AP}$. From this mode the only possible transition to a
full-duplex mode is $\mathsf{AP}\leftrightarrows \mathsf{M}_1$. The
mechanism of two-way setup is discussed in Section~\ref{sec:srb}. The
FD-MAC protocol therefore allows all modes to occur by switching
between various modes through mechanisms introduced by FD-MAC, and
some existing 802.11 capability.


\begin{figure}[ht]
\centerline{\resizebox{2.5in}{!}{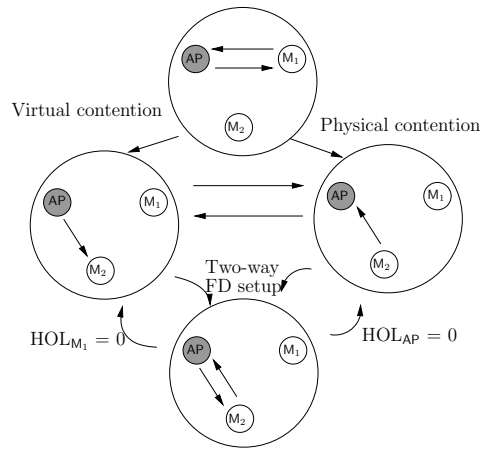}}
\caption{Switching between different modes of operation in a clique
  topology. The part of the state diagram illustrating all the key
  features of the FD-MAC is shown. State diagram has two more
  half-duplex modes $\mathsf{AP} \rightarrow \mathsf{M}_1$ and
  $\mathsf{M}_1 \rightarrow \mathsf{AP}$}
\label{fig:states}

\end{figure}

The hidden node topology with two mobile nodes and an $\mathsf{AP}$ is
shown in Figure~\ref{fig:FD-capable}(b). With $\mathsf{M}_1$ and
$\mathsf{M}_2$ hidden with respect to each other, two full-duplex
flows in addition to $\mathsf{AP} \leftrightarrows \mathsf{M}_1$,
$\mathsf{AP} \leftrightarrows \mathsf{M}_2$ are possible. They are
$\{\mathsf{M}_2 \rightarrow \mathsf{AP}, \mathsf{AP}\rightarrow
\mathsf{M}_1 \}$, and $\{\mathsf{AP} \rightarrow \mathsf{M}_2,
\mathsf{M}_1 \rightarrow \mathsf{AP}\}$. Each of four full-duplex
modes, whether two-way exchange or otherwise start with a particular
half-duplex mode. For instance $\{\mathsf{M}_2 \rightarrow
\mathsf{AP}, \mathsf{AP}\rightarrow \mathsf{M}_1 \}$ is possible only
if there exists $\mathsf{AP}\rightarrow \mathsf{M}_1$ as discussed in
Section~\ref{sec:snoop}. 

In order to ensure that transition to all full-duplex modes is
possible, FD-MAC must ensure that the half-duplex mode needed to kick
start it is possible. Half-duplex modes among themselves contend via
802.11 type of physical contention. The two-way full-duplex exchanges
have a period of shared random backoff for other half-duplex modes to
occur. Moreover they also have virtual contention resolution at the
$\mathsf{AP}$ to allow different half-duplex modes. On the other hand
$\mathsf{AP} \leftrightarrows \mathsf{M}_2$, $\{\mathsf{M}_2
\rightarrow \mathsf{AP}, \mathsf{AP}\rightarrow \mathsf{M}_1 \}$ type
of full-duplex, has $\mathsf{AP}$ always contending for the media
after it finishes sending the ACK, thus allowing all other nodes to
contend and establish a half-duplex communication with $\mathsf{AP}$.
Since all half-duplex modes are possible from any starting
state. Consequently the snooping mechanism will allow all full-duplex
modes too.


\vspace{-0.2cm}
\subsection{FD-MAC evaluations on WARP}
\label{sec:fd-maceval}
In this section we evaluate the FD-MAC for a two node full-duplex
exchange by implementing it on a real time
full-duplex system designed using WARP. Figure~\ref{fig:warp} shows a
full-duplex WARP node, with one transmit and one receive antenna.

\begin{figure}[ht]
\centering
  \includegraphics[width=0.22\textwidth]{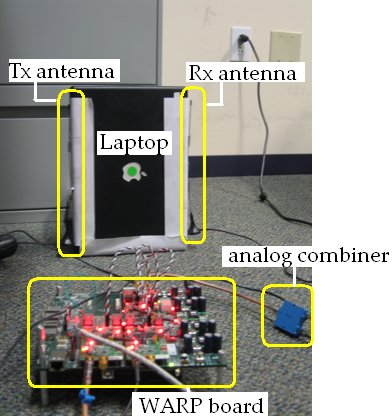}
 \caption{A full-duplex WARP node}
 \label{fig:warp}
\vspace{-0.3cm}
\end{figure}

The experimental set-up has two full-duplex nodes exchanging packets
with each other. FD-MAC ensures setting up of the full-duplex upon
discovering an opportunity to exchange packets in full-duplex
mode. The buffer at both the nodes always had a head of line packet
for the other. The evaluation compares the throughput of full-duplex
against half-duplex (again implemented on WARP).
\begin{table}[h]
\centering
\vspace{-0.2cm}
\caption{Number of packets/sec}
\begin{tabular}{|c|c|c|}\hline
SINR & Throughput & Throughput \\
of FD & of FD & HD \\ \hline
9dB & 285 & 158 \\ \hline
8dB & 276 & 165 \\ \hline
7dB & 253 & 169 \\ \hline
5dB & 269 & 151 \\ \hline
\end{tabular}
\vspace{-0.3cm}
\end{table}
 The modulation used for transmission was QPSK.

\emph{Result 7 (Increase in throughput due to full-duplex)}: The
encouraging result is that the throughput of full-duplex two-way
exchange using FD-MAC is 70\% higher than that of half-duplex for
identical transmit power.

\vspace{-0.2cm}



\section{Discussion and Conclusions}
\label{sec:conclude}

We note that FD-PHY and FD-MAC are first real-time design and
implementation of full-duplex physical and medium access layers, and
thus expect many avenues to further optimize the system
performance. Perhaps the most promising is a joint design of
transmitted signal, canceling mechanisms and baseband processing on
full-duplex nodes. We believe that this could lead to further
self-interference suppression, and perhaps push the performance to
near-perfect full-duplex systems.

Considering that full-duplex is still viewed skeptically by many, it
is crucial to demonstrate real-time implementations showing fully
operational network stacks. Towards that end, our work shows very
promising results, creating a strong case for practical use of
full-duplex in deployed networks.


\bibliographystyle{unsrt}
\scriptsize
\bibliography{FDbib}  

\end{document}